\documentclass[fleqn,10pt]{wlscirep}
\usepackage[utf8]{inputenc}
\usepackage[T1]{fontenc}
\usepackage[utf8]{inputenc}
\usepackage[T1]{fontenc}
\usepackage{lineno,soul,color}
\usepackage{longtable}
\usepackage{graphicx}
\usepackage{tabularx,booktabs}
\usepackage{url}
\usepackage[figuresright]{rotating}
\usepackage{fancybox}
\usepackage{dirtytalk}
\usepackage[utf8]{inputenc}
\usepackage{array}
\usepackage{adjustbox}
\usepackage{tabulary}
\usepackage{xcolor,colortbl}

\title{Digitally-Enhanced Dog Behavioral Testing: Getting Help from the Machine}

\author[1]{Nareed Farhat}
\author[2]{Teddy Lazebnik}
\author[3]{Joke Monteny}
\author[3]{Christel Palmyre Henri Moons}
\author[3]{Eline Wydooghe}
\author[4]{Dirk van der Linden}
\author[1,+]{Anna Zamansky}

\affil[1]{University of Haifa, Haifa, Israel}
\affil[2]{University College London, London, UK}
\affil[3]{VIVES University College, Roeselare, Belgium}
\affil[4]{Northumbria University, Newcastle upon Tyne, UK}

\affil[*]{annazam@is.haifa.ac.il, lazebnik.teddy@gmail.com}

\begin{abstract}
The assessment of behavioral traits in dogs is a well-studied challenge due to its many practical applications such as selection for breeding, prediction of working aptitude, chances of being adopted, etc. Most methods for assessing behavioral traits are questionnaire or observation-based, which require a significant amount of time, effort and expertise. In addition, these methods are also susceptible to subjectivity and bias, making them less reliable. In this study, we proposed an automated computational approach that may provide a more objective, robust and resource-efficient alternative to current solutions. Using  part of a \say{Stranger Test} protocol, we tested \(n=53\) dogs for their response to the presence and benign actions of a stranger. Dog coping styles were scored by three experts. Moreover, data were collected from their handlers using the Canine Behavioral Assessment and Research Questionnaire (C-BARQ) . An unsupervised clustering of the dogs' trajectories revealed two main clusters showing a significant difference in the stranger-directed fear C-BARQ factor, as well as a good separation between (sufficiently) relaxed dogs and dogs with excessive behaviors towards strangers based on expert scoring. Based on the clustering, we obtained a machine learning classifier for expert scoring of coping styles towards strangers, which reached an accuracy of 78\%. We also obtained a regression model predicting C-BARQ factor scores with varying performance, the best being Owner-Directed Aggression (with a mean average error of 0.108) and Excitability (with a mean square error of 0.032). This case study demonstrates a novel paradigm of digitally enhanced canine behavioral testing.  
\end{abstract}
\begin{document}

\flushbottom
\maketitle
%
%
\thispagestyle{empty}

\section*{Introduction}
Behavioral traits in animals are consistent patterns of behaviors exhibited in similar situations \cite{ilska2017genetic,dowling2011behavioral}. They are driven by personality \cite{svartberg2007individual}, which is a complex combination of genetic, cognitive, and environmental factors \cite{krueger2008behavioral}. The assessment of personality traits in dogs is gaining increasing attention due to its many practical applications in applied behavior \cite{arata2010important,sinn2010personality,scarlett2007aggressive,maejima2007traits}. Some examples of such applications include determining the suitability of working dogs \cite{wilsson1997use}, identifying problematic behaviors \cite{netto1997behavioural}, and adoption-related issues for shelter dogs \cite{dowling2011behavioral}. 

Measuring behavioral traits of dogs has been an enigmatic challenge in scientific literature for decades. The methods used can be roughly divided into two types. The first type refers to experimental behavioral tests (e.g., observations of the dog’s behavior in a controlled novel situation, such as the Strange Situation Test \cite{palestrini2005heart}). Brady et al.\cite{brady2018systematic} provide a systematic review of the reliability and validity of behavioral tests that assess behavioral characteristics important in working dogs. Jones and Gosling \cite{jones2005temperament} provide another comprehensive review of past research on canine temperament and personality traits. In a complementary manner, Bray et al.\cite{bray2021enhancing} reviewed 33 empirical studies assessing the behavior of working dogs. Tests for detection dogs have also been addressed\cite{la2017identifying,troisi2019behavioral,lazarowski2020selecting}. 
Another common assessment method is a questionnaire completed by the owner or handler .  Examples include the Monash Canine Personality Questionnaire \cite{ley2008personality}, the Dog Personality Questionnaire \cite{mirko2012preliminary}, VIDOPET \cite{turcsan2018personality} and many more. One of the most well-known questionnaires, used in many contexts, is the Canine Behavioral Assessment and Research Questionnaire (C-BARQ). Originally developed in English\cite{hsu2010factors,serpell2001development}, it has been validated in a number of languages, including Dutch \cite{van2006phenotyping}.

Although questionnaires can reduce the time and effort required for behavioral testing, they have serious limitations: they are susceptible to subjectivity and misinterpretation, and can be biased by the bond with the animal being assessed. Moreover, an individual with sufficient knowledge of the dog in order to reliably complete the items is not always available \cite{brady2018systematic}, especially in the case of working or shelter dogs. 

For instance, in the context of owner-observed assessment of stress, Mariti et al. have argued that ``The results show that some owners can help in protecting the welfare of their dogs, but that many owners would benefit from educational efforts to improve their ability to interpret their dogs’ behavior." \cite{mariti2012perception}. Moreover, Mariti et al. have shown that ``most dog owners report having a good understanding of the emotional state of their dogs, when they seem to have a low appreciation of the signals that dogs send in the earlier stages of emotional arousal" \cite{kerswell2009self}. Even seemingly clear physical observations, such as obesity in dogs, have been shown to lead to frequent disagreements between owners and veterinarians \cite{white2011canine}. Furthermore, Rayment et al. \cite{rayment2015applied} criticize the ``lack of robust assessment of the validity and reliability of many test protocols currently in use ‘on the ground’", referring in particular to the use of psychometric instruments that rely on an unambiguous shared understanding of terminology, which is difficult to achieve in a population with different levels of education, knowledge of animal behavior, (first) languages, etc. Moreover, psychological factors of the human observers influence their evaluation of dogs \cite{kujala2017human}, which further complicates the use of psychometric data from a wide variety of participants as a homogenous dataset of observations.

The goal of this exploratory study is to investigate a novel idea of a {\em digital enhancement} for behavioral testing, which in time may be integrated into relevant interspecies information systems~\cite{van2021interspecies} to understand animal behavior. Using as a case study a simple behavioral testing protocol of coping with the presence of a stranger, currently implemented to improve breeding of working dogs in Belgium, we ask the following questions: 

\begin{itemize}
\item Can machine identify different `behavioral profiles' in an objective, `human-free' way, and how do these profiles relate to the scoring of human experts in this test?  

\item Can machine predict scoring of human experts in this test? 

\item Can machine predict C-BARQ factors of the participating dogs? 
\end{itemize}

The findings of our study provide positive answers to these questions. Dog trajectories obtained by automated tracking of dog movements were clustered using an unsupervised k-mean clustering algorithm, revealing two distinct clusters corresponding to different human expert scoring categories (neutral vs. excessive behavior towards a stranger). Based on the clustering, we obtain a machine learning classifier that predicts expert scores with over 78\% Accuracy. We further obtain a regression model that predicts C-BARQ factor scores with varying performance, the best being Owner-Directed Aggression (with a mean average error of 0.108) and Excitability (with a mean square error of 0.032). We discuss the potential applications of the proposed novel paradigm of digitally enhanced canine behavioral testing.

\section*{Methods}
{\em Ethical statement.} All experiments were performed in accordance with relevant guidelines and regulations. The experimental procedures and protocols were reviewed by the Ethical Committees of KU Leuven and University of Haifa, in both ethical approval was waived. 

\begin{figure}[htb!]
\centering
\includegraphics[width=.8\columnwidth]{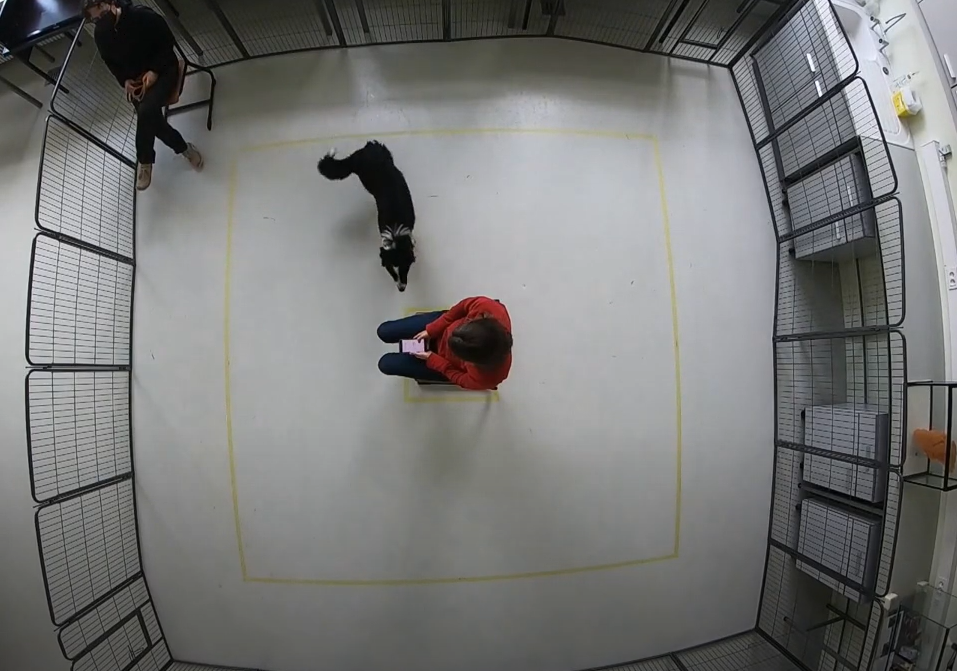}
\caption{The testing arena, captured from the top camera; stranger sitting in the middle, handler in the corner. }
\label{fig:arena}
\end{figure}

\subsection*{Testing Arena}

The test was conducted indoors, in a room free from other distractions such as other animals or people (except the test person, the assistant, and the familiar person/owner). The size of the test arena was 6*8 m (length x width) with a height of 3 m. An adjacent, separate room is available where the dog and the owner are received and can wait out of sight of the test arena. The room containing the test arena has a separate door, so that the test person can enter the test arena without having to pass by the dog and owner. In this way, the test person is completely new to the dog until the start of the actual test. 
The test arena was enclosed with metal fences (height of 0.8 m, length x width of 4.7 by 4.7 m). In the middle of the test arena, a square of 60 by 60 centimeters is drawn with tape for positioning the chair of the test person. The test person faces the front of the fenced arena. This is the side that contained the gate and from where the assistant recorded the test on video. In the left corner (frontal view), there is a chair for the familiar person, positioned parallel to the front fence.  A second square (3 x 3 m), centered around the middle, was marked with tape on the floor. These lines indicated the track to be followed when the owner or test person walked in the test arena. 

Two video cameras were used to record the activity and the behavior of the dog during the test, a top view and a side view camera. As top view, a GoPro video camera was mounted in the middle of the test arena at a height of approximately 3 m, so that the entire test arena was covered - see Figure \ref{fig:arena}. A side-view camera was held and operated by the assistant, at a distance between 1.5 and 2 m from the middle of the nearest fence.

\subsection*{Test Procedure}
The protocol used below is part of a more elaborate testing protocol developed by one of the authors (JM). Its purpose is to assess the reactions of dogs to the presence of an unfamiliar person (during inactivity or during benign actions), both in the presence and absence of a familiar person (i.e., the owner or a regular handler/trainer). The familiar person is instructed about the test and asked not to interact with the dog, expect to unleash or put the leash on the dog. 
For the purposes of this study, only the exploration phase and the first phase occurred. The unfamiliar person, i.e., the test person, was always the same adult female (JM). The assistant was also always an adult female, but not always the same person.

\subsubsection*{Exploration Phase}

Prior to the first phase of the behavioral test, all dogs were allowed to explore the test arena in the presence of their owner, but without the presence of the test person.  An assistant led the familiar person into the test arena with the dog on a leash. While the assistant moved to the location from where the next phase would be filmed, the familiar person entered the test arena with the dog, closed the gate and unleashed the dog. The familiar person then walked the 3m x 3m square counterclockwise 1.75 times, arriving at and sitting on the chair positioned on the side. After three minutes, the familiar person walked to the dog and put on the leash. They both left the test arena and went into the adjacent room.

\subsubsection*{Testing Phase}

After the exploration phase, the test person entered the test arena and sat on the chair in the middle, feet in parallel and firmly planted. The test person held a smartphone as a timer. 
The assistant called in the familiar person and the dog. They entered the test arena, the familiar person closed the gate, unleashed the dog, walked directly to the chair and sat down. 
After the familiar person sat down, the test person performs three actions: a short, clear cough (at 10 s), a hand running through the hair for 3 seconds (at 20 s), and crossing the right leg over the left (at 30 s). These are actions that can be expected from any human being and that all dogs will encounter when they are around people. An example trial can be found \href{https://drive.google.com/file/d/1VpaxKePw2ICY2SMGPwK_T4Td2zGyAOzc/view?usp=sharing}{here}.
Except when running her hand through her hair or when a dog jumps up, the test person held the smartphone in both hands, resting on her lap. The test person did not look at the dog or performed any actions towards it. If a dog jumped up excitedly, the test person protected her face/head with her hands/arms as needed. In this study we analyze this testing phase.

\subsubsection*{Study Subjects}
A total of n=53 dogs were tested in the study. Their owners were recruited through social media in Belgium. The inclusion criteria for the dogs were: 
\begin{itemize}
    \item Age: between 1 and 2 years old. 
    \item Height: between 30 and 65 centimeters.
    \item Up-to-date vaccinations and no history of health problems.
    \item Accompanied by an owner or a familiar person. 
    \item Belonging to the modern dog breeds. 
\end{itemize}

\subsubsection*{Dog Scoring}
\label{sec:scoring}

The original scoring method, an 11-point scale ranging from -5 to +5 was developed by JM. It is based on an adaptation of the concept of coping with potential threats via freeze/flight versus fight \cite{Riemer2013ChoiceOC}. The idea is to differentiate between two main tendencies of dogs when reacting to a stressor (in this context - an unfamiliar person – reactions to the assistant or familiar person were ignored): dogs that tend to ‘react towards the stressor’ (e.g. get very close to the test person, jump up, chew, show offensive aggression…) receive numerically positive scores, and dogs that tend to ‘react away from the stressor’ (e.g. keep at a distance, avoid, show defensive aggression…) receive numerically negative scores. Score 0 (neutral) indicates good and stable coping with the stressor. 

The original scoring method is currently used by the Belgian assistance dog breeding organization, Purpose Dogs vzw,\footnote{https://purpose-dogs.be/} to improve breeding outcomes. In order to facilitate the scoring by the independent experts, the original scale ranging from -5 to +5
was modified into a five-point scale ranging from -2 to +2 (-5 and -4 were scored as -2; -3 and -2 as -1; -1, 0 and +1 as 0; +2 and +3 as +1; +4 and +5 as +2). Since this category includes as well the -1 and +1 on the 11-point scale, a score of 0 on the five-point scale means that the dog is sufficiently relaxed in the presence of the test person. In the case of Purpose Dogs vzw, for example, dogs with this score would be suitable for selection for future breeding.
The analysis for the purpose of this study was further simplified by grouping the negative (-2 and -1) and positive (+2 and +1) scores, respectively, resulting in three groups: “+”, “0”, and “-“.

The testing phase (phase 1) was evaluated by three experts. Two of the experts were animal behavior researchers (JM and CPHM), and one expert was a veterinarian (EW). Multi-rater (Fleiss) kappa on the scores (n=53) collapsed into three classes (`-', 0, '+') reached a percentage of agreement of 85\%; Fleiss free-marginal k=0.77 indicating good strength of inter-rater reliability. To obtain an overall score for each dog, the score representing the majority (two out of three scores) was selected as the final score. Negative scores had only 3 samples, so this category was excluded from our analysis. Our final dataset included 50 samples, out of them 32 samples with a zero/neutral score (26 full agreement by all coders, 6 by majority) and 18 samples with a positive/excessive score (12 full agreement by all coders, 6 by majority).

\subsubsection*{C-BARQ Questionnaire}\label{sec:factors}

The Canine Behavioral Assessment and Research Questionnaire (C-BARQ) is an instrument originally developed in English \cite{hsu2010factors,serpell2001development} and validated in multiple languages, including Dutch \cite{van2006phenotyping}.

In the context of our study, we used the following eight factors identified in Hsu et al. \cite{hsu2010factors} using factor analysis: (1) Stranger directed aggression (SDA), (2)	Owner directed aggression (ODA), (3) Stranger directed fear (SDF), (4) Nonsocial fear (NSF), (5)	Separation related behavior (SRB), (6)	Attachment seeking behavior (ASB), (7)	Excitability (EXC), and (8)	Pain sensitivity (PS). 

The dog owners were asked to complete a Dutch\footnote{The C-BARQ was translated into Dutch using a two-step translation approach with bilingual speakers.} version of the C-BARQ questionnaire.

\subsection*{Computational Approach}

\begin{figure}[htb!]
\centering
\includegraphics[width=0.9\columnwidth]{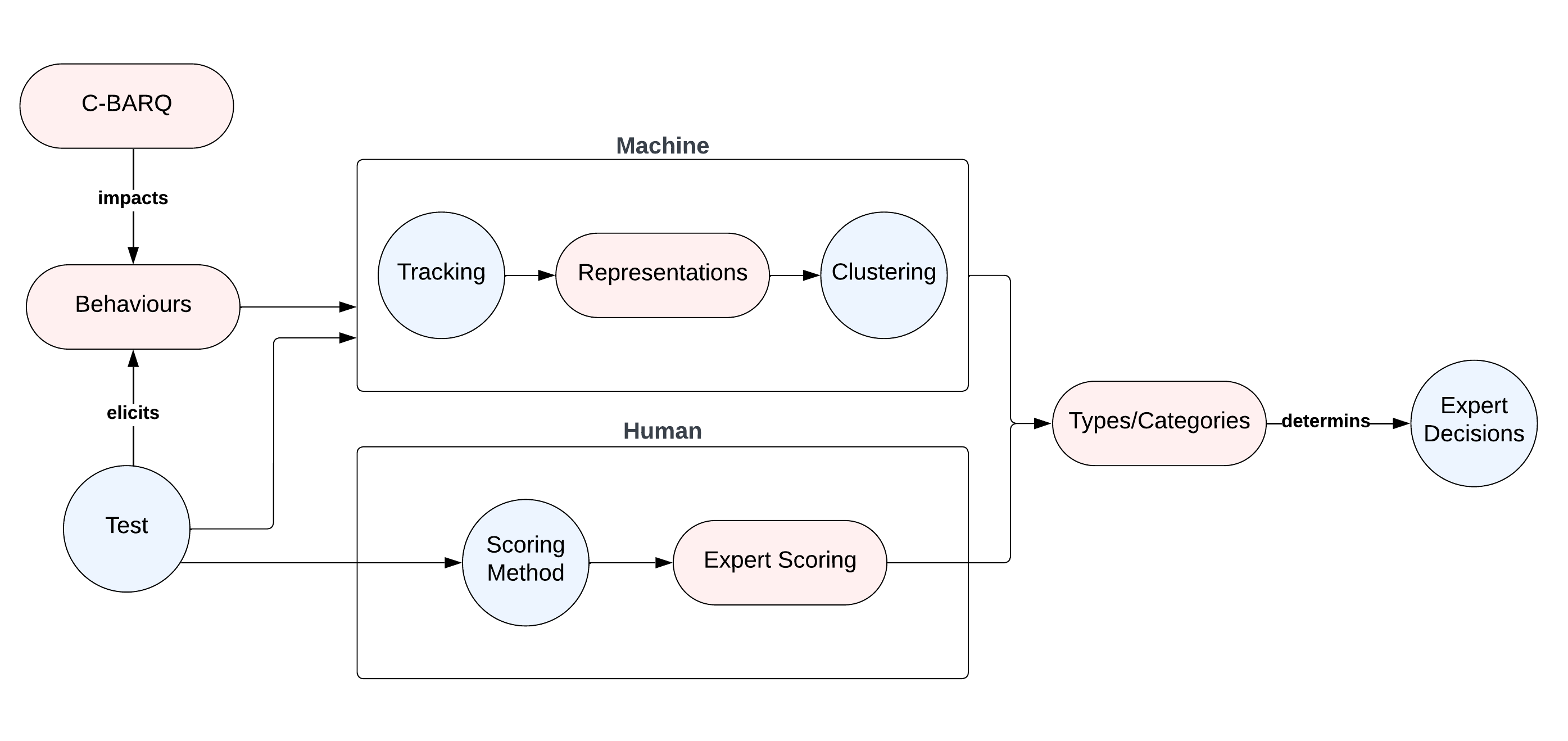}
\caption{A conceptual framework for digitally enhanced dog behavioral testing}
\label{fig:frm}
\end{figure}

Figure \ref{fig:frm} provides an overview of our approach for digital enhancement of dog behavioral testing. The purpose of any behavioral test is, eventually, to observe behaviors in response to various stimuli in a controlled and standardized environment. Based on a specific testing protocol, a scoring method is usually developed and evaluated for use by human experts. The practical aim of such scoring is to classify the elicited behaviors into categories (e.g. corresponding to specific behavioral traits or profiles) that can eventually be used for decision support. With the machine entering the scene, we have an alternative, mathematical and {\em completely human-free} way of ``scoring" behaviors, or dividing them into categories. Since this test focuses on human-directed behavior, we assume that the participants' trajectories contain meaningful behavioral information about their reaction to the stranger. Therefore, we automatically extract and cluster the participants' trajectories, investigating the relationship of the emerging clusters to experts' scoring, and compare how well they align. In addition, we also examine the correlation of the emerging clusters with C-BARQ factors. Finally, we investigate whether we can use the obtained clusters to predict (i) the expert score, and (ii) C-BARQ factors. 

In what follows we provide further details on the tracking method, the clustering method and the machine learning models for prediction of the above.

\subsubsection*{Tracking Method}
The BLYZER system is a self-developed platform that
aims to provide a flexible automated behavior analysis which has been applied in several studies for analyzing dog behavior\cite{karl2020exploring,zamansky2019analysis,bleuer2019computational,fux2021objective}. A similar approach was implemented on a smaller portion of the dataset used in this study in \cite{menaker2022clustering}, however in contrast to our approach here, features chosen manually were used for clustering. 

\begin{figure}[htb!]
\centering
\includegraphics[width=.7\columnwidth]{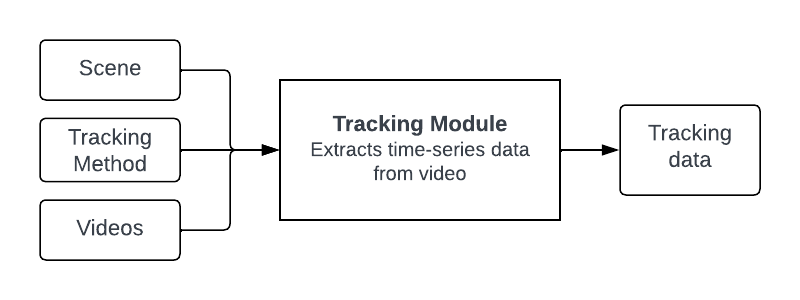}
\caption{BLYZER tracking module architecture.}
\label{fig:blyzerarch}
\end{figure}

\begin{figure}[htb!]
\centering
\includegraphics[width=.9\columnwidth]{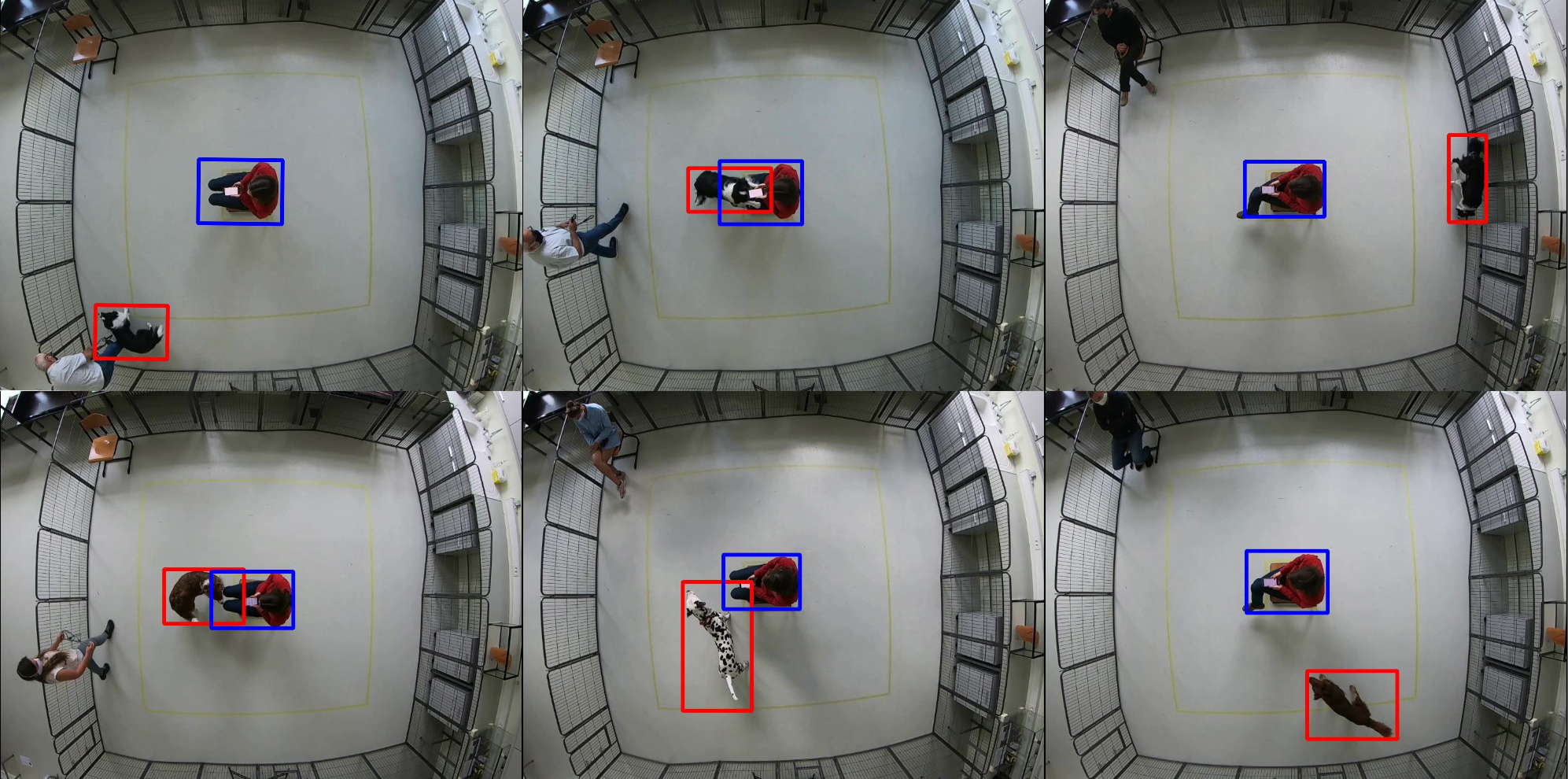}
\caption{Frames example of participating dog and test person being tracked by Blyzer.}
\label{fig:detectioneg}
\end{figure}

\begin{figure}[htb!]
\centering
\includegraphics[width=.9\columnwidth]{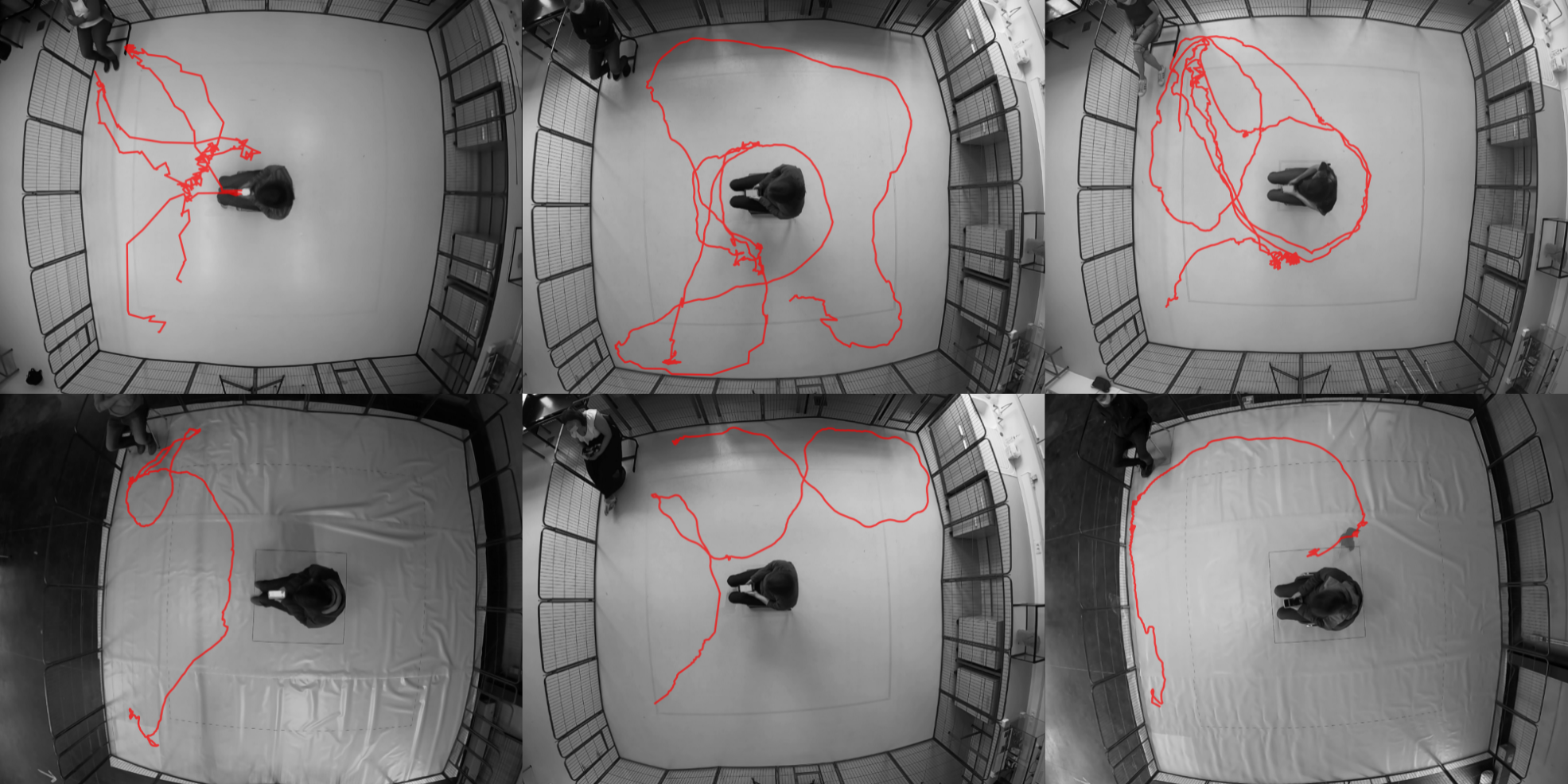}
\caption{Examples of participating dogs' trajectories extracted with BLYZER; top: scoring `+', bottom: scoring 0. }
\label{fig:traj}
\end{figure}

BLYZER's input is video footage of a dog freely moving in a room and possibly interacting with objects, humans or other animals, while its output is time series represented in a json file with the detected locations of the objects in each frame. Figure \ref{fig:blyzerarch} shows the pipeline, highlighting the fact that both tracking method (the models used for detection) and the scene (amount of moving and fixed objects) are easily adapted. In the configuration used in this paper, the tracking method was chosen to be a neural network based on the Faster R-CNN architecture \cite{Ren2015FasterRT}
pre-trained on the COCO 2017 dataset \cite{Lin2014MicrosoftCC}, which we retrained on additional 106,768 images of two objects: a person and a dog. The images were collected from (1) Open image dataset V6 \cite{Kuznetsova2018TheOI} (2) Pascalvoc dataset \cite{Everingham2010ThePV} (3) COCO dataset \cite{Lin2014MicrosoftCC} (4) Images from previous studies\cite{bleuer2019computational,fux2021objective}. Figure \ref{fig:detectioneg} shows example frames from our dataset with dog and test person object detection. And figure \ref{fig:traj} presents examples of participants' trajectories extracted with BLYZER.

{\em Quality of detection.} To ensure sufficient tracking, only videos with a percentage of frames where dog and person are correctly detected of least 80\% of the frames, leading to the exclusion of three videos (all three scored with a zero/neutral score). For the remaining 47 videos, we applied post-processing operations available in BLYZER to remove noise and enhance detection quality using smoothing and extrapolation techniques for the dog and test person detection, reaching almost perfect (above 95\%) detection.  

\subsection*{Clustering Method}

\begin{figure}[htb!]
\centering
\includegraphics[width=1\columnwidth]{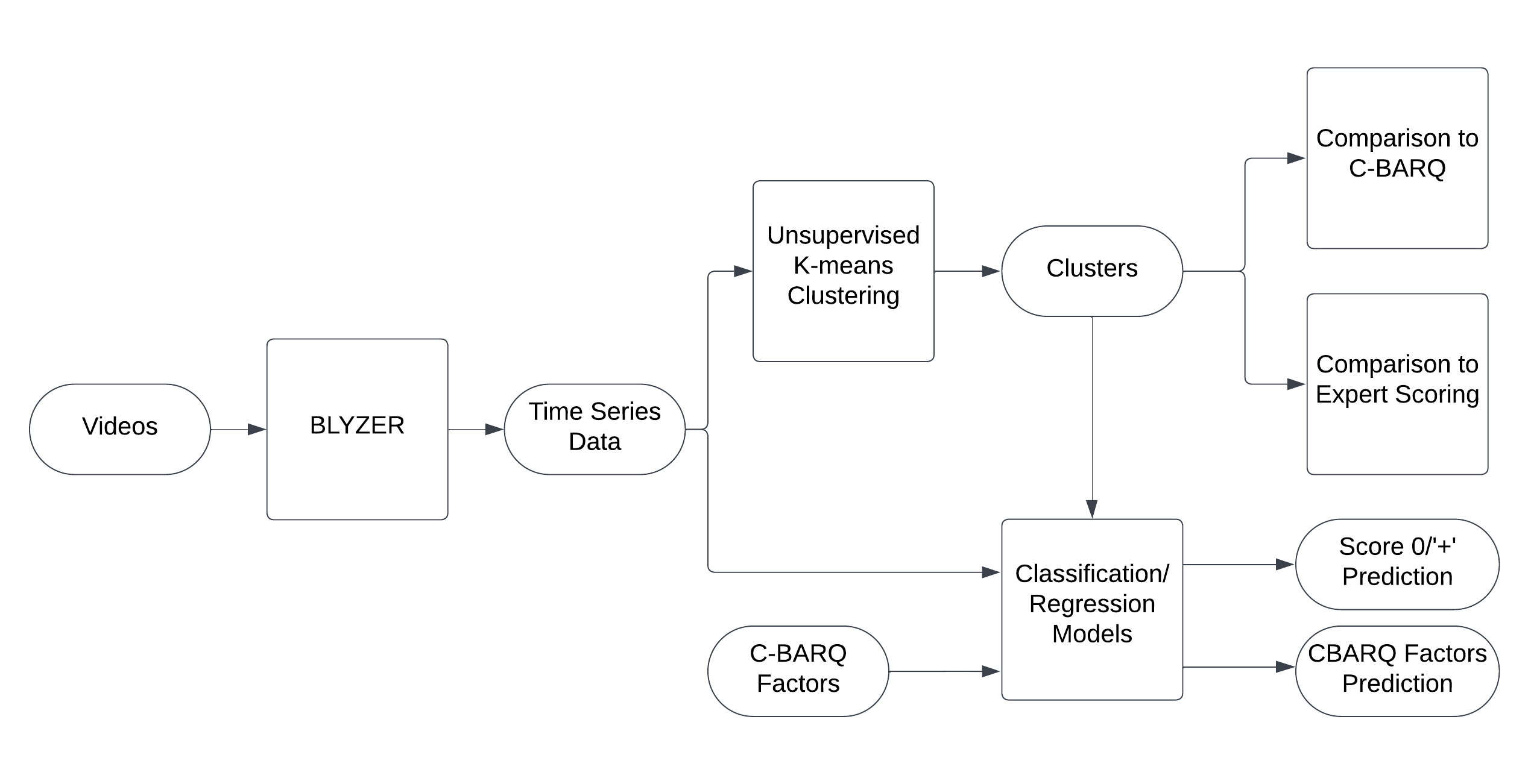}
\caption{Data Analysis Pipeline.}
\label{fig:clustering}
\end{figure}

The videos from the trails are initially analyzed by the BLYZER tool which produces for each frame the center of mass of the dog and person in the frame (if detected). To assure a smooth motion capture while standardizing between trials, we set 24 frames per second (FPS) rate across all videos. For frames that the BLYZER tool was not able to detect either the dog or the person (or both), it linearly extrapolates their positions to fulfill the gap. In addition, since not all videos were of identical duration, we used the duration of the shortest video as standard duration. As such, each trial (\(s \in \mathbb{R}^{2m}\)) is defined by a time series with a fixed duration between samples constructed by two vectors, one for the dog's position \((d \in \mathbb{R}^m)\) and the other for the person's position \((p \in \mathbb{R}^m)\). As a result, we obtain a dataset, \(D \in \mathbb{R}^{n \times m}\). This is the times series data depicted in Figure \ref{fig:clustering}, which presents the whole data analysis pipeline. 

For clustering trajectories, we used the time-series K-mean clustering algorithm \cite{ts_kmeans} with the elbow-point method \cite{elbow} to find the optimal number of clusters (\(k\)). Nonetheless, as the raw center of mass is not quite an informative space, we decided to first transform the data into a \say{movement} space. To this end, we trained a small-size one-dimensional convolutional neural network (CNN) based AutoEncoder model \cite{cnn_ae} with the following architecture for the encoder: Convolution with a window size of 3, dropout with \(p=0.1\),  Convolution with a window size of 3, dropout with \(p=0.1\), max-pooling with a window size of \(2\). Clearly, the decoder's architecture is opposite to the encoder's one. We used a mean absolute error as the metric for the optimization process and the ADAM optimizer \cite{adam}. The model's hyperparameters are found using a grid-search \cite{grid_search}. Using the encoder part of the model that was used after training the AutoEncoder, we computed the \say{movement} space of each sample for the clustering. 
Once the clustering is obtained, the clusters were evaluated in two ways: (i) expert scoring metrics, and (ii) Mann Whitney U test on C-BARQ factors.

\subsection*{Classification and regression machine learning models}
The clustering was further used to obtain classification and regression models for predicting scoring (0/'+') and C-BARQ factors respectively. 
%
We use the Tree-Based Pipeline Optimization Tool (TPOT), the genetic algorithm-based automatic machine learning library \cite{tpot}. TPOT produces a full machine learning (ML) pipeline, including feature selection engineering, model selection, model ensemble, and hyperparameter tuning; and shown to produce impressive results in a wide range of applications \cite{teddy_2,teddy_3,teddy_4}. Hence, for every configuration of source and target variables investigated, we used TPOT, allowing it to test up to \(10000\) ML pipelines. We choose \(10000\) to balance the ability of TPOT to converge into an optimal (or at least close to optimal) ML pipeline and the computational burden associated with this task.

The obtained classification model performance for expert scoring was evaluated using commonly used metrics of accuracy, precision, recall, and \(F_1\) score. The obtained regression model performance for C-BARQ factors was evaluated using Mean Absolute Error \cite{Frnkranz2010} (MAE), Mean Squared Error \cite{Frnkranz2010} (MSE), and R-squared \cite{Ling1981CorrelationAC} (\(R^2\)).

\section*{Results}

\subsection*{Clusters correlation with scoring}
Using the elbow method, two clusters emerged of sizes 26 and 20 respectively. One sample was excluded due to being an outlier. As shown in Table \ref{table:confusion}, there is a quite good separation between zero/neutral scores and positive/excessive scores:  the first cluster had the majority of participants (n=21) scoring 0, while only 5 scored '+'. The second, the majority (n=13) scored '+' while 7 scored 0. 

\begin{table}[!ht]
\centering
\begin{tabular}{l|cc|c}
                   & \textbf{Cluster 1  }      & \textbf{Cluster 2}         & \textbf{Total} \\ \hline
\textbf{Score 0}   & \cellcolor[HTML]{9AFF99}21& \cellcolor[HTML]{FFCCC9}7  & 28       \\
\textbf{Score '+'} & \cellcolor[HTML]{FFCCC9}5 & \cellcolor[HTML]{9AFF99}13 & 18      \\ \hline
\textbf{Total}     & 26                        & 20                         & 46           
\end{tabular}
\caption{Cluster description in correlation with expert scoring.}
\label{table:confusion}
\end{table}

\begin{table}[!ht]
\centering
\begin{tabular}{lcccc}
\multicolumn{1}{l|}{} &  \multicolumn{1}{c|}{Precision} &
  \multicolumn{1}{c|}{Recall} & \multicolumn{1}{c|}{F1} & Accuracy \\ \hline

\multicolumn{1}{l|}{\textbf{Score Classifier}} & \multicolumn{1}{c|}{\textbf{0.771}} &
  \multicolumn{1}{c|}{\textbf{0.782}} & \multicolumn{1}{c|}{\textbf{0.775}} & \textbf{0.787}
\end{tabular}
\caption{Evaluation metrics.}
\label{table:metrics}
\end{table}

\subsection*{Clusters correlation with C-BARQ}





A significant difference between the two clusters was found with respect to Stranger-Directed Fear (SDF) (median of cluster 1: 0.00; cluster 2: 0.42) (Mann Whitney U=120.5, z=-2.56, p=0.01). No other factors of the C-BARQ had a significant difference with respect to the clusters. 

\subsection*{Performance of Expert Score Classification Model}

Table \ref{table:metrics} presents the performance of the expert score classification model, reaching accuracy of above 78\%. 


\subsection*{Performance of C-BARQ Factors Regression Model}

Our findings revealed varying levels of error across the eight C-BARQ factors presented above. Owner directed aggression (ODA) yields the lowest MAE of 0.108 and second lowest MSE of 0.046. Excitability (EXC) has the second-lowest MAE of 0.122, and the lowest MSE of 0.032. Stranger directed aggression (SDA) and Separation related behavior (SRB) demonstrated slightly higher MAE values of 0.275 and 0.257, respectively, with corresponding MSE values of 0.129 and 0.144. Moreover, Pain sensitivity (PS), Stranger directed fear (SDF), Non social fear (NSF), and Attachment seeking behavior (ASB) exhibited MAE values ranging from 0.435 to 0.51, with MSE values ranging from 0.287 to 0.43. Furthermore, the \(R^2\) values provided insights into the proportion of variance explained by the factors. EXC showed the highest \(R^2\) value of 0.886, indicating a strong fit between the EXC factor and the time-series data. SDA and PS exhibited moderate \(R^2\) values of 0.47 and 0.429, respectively, signifying a reasonable level of explanatory power. Moreover, ODA and ASB demonstrated low \(R^2\) values of 0.176 and 0.142. While SRB, NSF, and SDF showed the lowest \(R^2\) values of 0.073, 0.043, and 0.032, indicating a weak association with the time-series data. These outcomes shed light on the predictive performance of the model and the varying impacts of the C-BARQ factors on the outcome.

\begin{table}[!ht]
\centering
\begin{tabular}{l|l|l|l}
                                   & MAE   & MSE   & \(R^2\)    \\ \hline
Owner Directed Aggression (ODA)    & \textbf{0.108}& 0.046 & 0.176 \\ \hline
Excitability (EXC)                 & 0.122 & \textbf{0.032} & \textbf{0.886} \\ \hline
Separation related behavior (SRB)  & 0.257 & 0.144 & 0.073 \\ \hline
Stranger directed aggression (SDA) & 0.275 & 0.129 & 0.47  \\ \hline
Pain sensitivity (PS)              & 0.435 & 0.319 & 0.429 \\ \hline
Non social fear (NSF)              & 0.438 & 0.334 & 0.043 \\ \hline
Attachment seeking behavior (ASB)  & 0.441 & 0.287 & 0.142 \\ \hline
Stranger directed fear (SDF)       & 0.510  & 0.430  & 0.032
\end{tabular}
\caption{Regression model metrics per C-BARQ factor, sorted by MAE.}
\label{table:regmetrics}
\end{table}

\section*{Discussion}
This study is another contribution to the growing field of computer-aided solutions for \say{soft} questions using data-driven based methods \cite{teddy_1,comp_help_1,comp_help_2,comp_help_3}. To the best of our knowledge, this study is the first to provide a machine-learning model for objectively scoring a strictly controlled dog behavioral test.

In this study we used as a case study a Stranger Test routinely performed in a working dog organization to ask the following questions: 

\begin{itemize}
\item Can machine identify different `behavioral profiles' in an objective, `human-free' way, and how do these profiles relate to the scoring of human experts in this test?  

\item Can machine predict scoring of human experts in this test? 

\item Can machine predict C-BARQ factors of the participating dogs? 
\end{itemize}

Our results indicate positive answers to all of the above questions. Answering the first question, using unsupervised clustering, two clusters emerged, with a good separation between the two different scores 0/`+'. Answering the second question, we presented a classification model for predicting human scoring reaching 78\% accuracy. Answering the third question, we presented a regression model which is able to predict Excitability and Owner-Directed Aggression with minimal error. These results demonstrate the potential of computational approaches in the domain of dog behavioral testing. 

It is important to stress that the computational approach to the assessment of dog behavioral testing proposed here is `human-free'. The agenda for a `human-free' computational analysis of animal behavior was introduced in Forkosh\cite{forkosh2021animal}. The author argued that despite the fact that automated tracking of animal movement is well-developed, the interpretation of animal behaviors remains human-dependent and thus inherently anthropomorphic and susceptible to bias. Indeed, in previous works applying computational approaches in the context of dog behavior \cite{fux2021objective,menaker2020towards,karl2020exploring,volter2023using}, features used for machine learning are explicitly selected by human experts. 

By using such ``human-free" clustering, two clusters emerged, roughly dividing the participants into a cluster of `neutrally reacting' dogs with the majority scoring 0, and a cluster with a majority of `excessively reacting' dogs scoring `+'. Interestingly, these clusters showed a significant difference in the Stranger-Directed Fear C-BARQ factor. However, a regression model for predicting this factor did not have a very good performance, with the best performance being the Owner-Directed Aggression factor and Excitability. The latter could be related to the excessive behaviors typical of the `+' scoring that matched the response of dogs as measured by the C-BARQ ``displaying strong reactions to potentially exciting or arousing events"\footnote{https://vetapps.vet.upenn.edu/cbarq/about.cfm}. Further research is needed to establish clearer relationships. 

The testing protocol used in our study refers to one specific aspect (towards/neutral/away from stressor) of stranger-directed behaviors. This protocol is used in a working dogs organization for breeding outcome improvement and has been previously studied in the context of automation of tracking\cite{menaker2020towards}, also exploring some preliminary ideas of clustering (unlike the `human-free' approach presented here). An in-depth exploration and scientific validation of this test is beyond the scope of the current study, we chose to use just one phase of this protocol due to its simplicity for automating tracking. Future research is needed for extending the presented approach to other phases of this protocol, as well as to other types of behavioral testing.  

Our approach in this study was validating the emerging clusters using expert scoring as a golden standard. However, this approach could be reversed in future studies, using mathematical, objective clustering as a `ground truth' for testing various scoring schemes for behavioral testing protocols. For now we treat the machine as enhancing human capabilities, however a day may come when this situation will be reversed, with the machine being the more objective and reliable way of analyzing behavioral testing data. It is our hope that this preliminary study will stimulate discussions on the value and great promise of AI in the context of dog behavioral testing.   

\section*{Acknowledgements}

The first and last authors were supported by Israel Ministry of Agriculture and Rural Development. 

\section*{Author contributions statement}

J.M., C.M., E.W. and A.Z. conceived the experiment(s),  T.L, D.L. and N.F. conducted the experiment(s). All authors analyzed the results and reviewed the manuscript. 

\bibliography{sample}

\begin{thebibliography}{10}
\urlstyle{rm}
\expandafter\ifx\csname url\endcsname\relax
  \def\url#1{\texttt{#1}}\fi
\expandafter\ifx\csname urlprefix\endcsname\relax\def\urlprefix{URL }\fi
\expandafter\ifx\csname doiprefix\endcsname\relax\def\doiprefix{DOI: }\fi
\providecommand{\bibinfo}[2]{#2}
\providecommand{\eprint}[2][]{\url{#2}}

\bibitem{ilska2017genetic}
\bibinfo{author}{Ilska, J.} \emph{et~al.}
\newblock \bibinfo{journal}{\bibinfo{title}{Genetic characterization of dog
  personality traits}}.
\newblock {\emph{\JournalTitle{Genetics}}} \textbf{\bibinfo{volume}{206}},
  \bibinfo{pages}{1101--1111} (\bibinfo{year}{2017}).

\bibitem{dowling2011behavioral}
\bibinfo{author}{Dowling-Guyer, S.}, \bibinfo{author}{Marder, A.} \&
  \bibinfo{author}{D’arpino, S.}
\newblock \bibinfo{journal}{\bibinfo{title}{Behavioral traits detected in
  shelter dogs by a behavior evaluation}}.
\newblock {\emph{\JournalTitle{Applied animal behaviour science}}}
  \textbf{\bibinfo{volume}{130}}, \bibinfo{pages}{107--114}
  (\bibinfo{year}{2011}).

\bibitem{svartberg2007individual}
\bibinfo{author}{Svartberg, K.}
\newblock \bibinfo{journal}{\bibinfo{title}{Individual differences in
  behaviour—dog personality}}.
\newblock {\emph{\JournalTitle{The behavioural biology of dogs}}}
  \bibinfo{pages}{182--206} (\bibinfo{year}{2007}).

\bibitem{krueger2008behavioral}
\bibinfo{author}{Krueger, R.~F.} \& \bibinfo{author}{Johnson, W.}
\newblock \emph{\bibinfo{title}{Behavioral genetics and personality: A new look
  at the integration of nature and nurture.}} (\bibinfo{publisher}{The Guilford
  Press}, \bibinfo{year}{2008}).

\bibitem{arata2010important}
\bibinfo{author}{Arata, S.}, \bibinfo{author}{Momozawa, Y.},
  \bibinfo{author}{Takeuchi, Y.} \& \bibinfo{author}{Mori, Y.}
\newblock \bibinfo{journal}{\bibinfo{title}{Important behavioral traits for
  predicting guide dog qualification}}.
\newblock {\emph{\JournalTitle{Journal of Veterinary Medical Science}}}
  \bibinfo{pages}{0912080094--0912080094} (\bibinfo{year}{2010}).

\bibitem{sinn2010personality}
\bibinfo{author}{Sinn, D.~L.}, \bibinfo{author}{Gosling, S.~D.} \&
  \bibinfo{author}{Hilliard, S.}
\newblock \bibinfo{journal}{\bibinfo{title}{Personality and performance in
  military working dogs: Reliability and predictive validity of behavioral
  tests}}.
\newblock {\emph{\JournalTitle{Applied Animal Behaviour Science}}}
  \textbf{\bibinfo{volume}{127}}, \bibinfo{pages}{51--65}
  (\bibinfo{year}{2010}).

\bibitem{scarlett2007aggressive}
\bibinfo{author}{Scarlett, J.}, \bibinfo{author}{Campagna, M.},
  \bibinfo{author}{Houpt, K.~A.} \emph{et~al.}
\newblock \bibinfo{journal}{\bibinfo{title}{Aggressive behavior in adopted dogs
  that passed a temperament test}}.
\newblock {\emph{\JournalTitle{Applied Animal Behaviour Science}}}
  \textbf{\bibinfo{volume}{106}}, \bibinfo{pages}{85--95}
  (\bibinfo{year}{2007}).

\bibitem{maejima2007traits}
\bibinfo{author}{Maejima, M.} \emph{et~al.}
\newblock \bibinfo{journal}{\bibinfo{title}{Traits and genotypes may predict
  the successful training of drug detection dogs}}.
\newblock {\emph{\JournalTitle{Applied Animal Behaviour Science}}}
  \textbf{\bibinfo{volume}{107}}, \bibinfo{pages}{287--298}
  (\bibinfo{year}{2007}).

\bibitem{wilsson1997use}
\bibinfo{author}{Wilsson, E.} \& \bibinfo{author}{Sundgren, P.-E.}
\newblock \bibinfo{journal}{\bibinfo{title}{The use of a behaviour test for the
  selection of dogs for service and breeding, i: Method of testing and
  evaluating test results in the adult dog, demands on different kinds of
  service dogs, sex and breed differences}}.
\newblock {\emph{\JournalTitle{Applied Animal Behaviour Science}}}
  \textbf{\bibinfo{volume}{53}}, \bibinfo{pages}{279--295}
  (\bibinfo{year}{1997}).

\bibitem{netto1997behavioural}
\bibinfo{author}{Netto, W.~J.} \& \bibinfo{author}{Planta, D.~J.}
\newblock \bibinfo{journal}{\bibinfo{title}{Behavioural testing for aggression
  in the domestic dog}}.
\newblock {\emph{\JournalTitle{Applied animal behaviour science}}}
  \textbf{\bibinfo{volume}{52}}, \bibinfo{pages}{243--263}
  (\bibinfo{year}{1997}).

\bibitem{palestrini2005heart}
\bibinfo{author}{Palestrini, C.}, \bibinfo{author}{Previde, E.~P.},
  \bibinfo{author}{Spiezio, C.} \& \bibinfo{author}{Verga, M.}
\newblock \bibinfo{journal}{\bibinfo{title}{Heart rate and behavioural
  responses of dogs in the ainsworth's strange situation: A pilot study}}.
\newblock {\emph{\JournalTitle{Applied Animal Behaviour Science}}}
  \textbf{\bibinfo{volume}{94}}, \bibinfo{pages}{75--88}
  (\bibinfo{year}{2005}).

\bibitem{brady2018systematic}
\bibinfo{author}{Brady, K.}, \bibinfo{author}{Cracknell, N.},
  \bibinfo{author}{Zulch, H.} \& \bibinfo{author}{Mills, D.~S.}
\newblock \bibinfo{journal}{\bibinfo{title}{A systematic review of the
  reliability and validity of behavioural tests used to assess behavioural
  characteristics important in working dogs}}.
\newblock {\emph{\JournalTitle{Frontiers in veterinary science}}}
  \textbf{\bibinfo{volume}{5}}, \bibinfo{pages}{103} (\bibinfo{year}{2018}).

\bibitem{jones2005temperament}
\bibinfo{author}{Jones, A.~C.} \& \bibinfo{author}{Gosling, S.~D.}
\newblock \bibinfo{journal}{\bibinfo{title}{Temperament and personality in dogs
  (canis familiaris): A review and evaluation of past research}}.
\newblock {\emph{\JournalTitle{Applied Animal Behaviour Science}}}
  \textbf{\bibinfo{volume}{95}}, \bibinfo{pages}{1--53} (\bibinfo{year}{2005}).

\bibitem{bray2021enhancing}
\bibinfo{author}{Bray, E.~E.} \emph{et~al.}
\newblock \bibinfo{journal}{\bibinfo{title}{Enhancing the selection and
  performance of working dogs}}.
\newblock {\emph{\JournalTitle{Frontiers in veterinary science}}}
  \bibinfo{pages}{430} (\bibinfo{year}{2021}).

\bibitem{la2017identifying}
\bibinfo{author}{La~Toya, J.~J.}, \bibinfo{author}{Baxter, G.~S.} \&
  \bibinfo{author}{Murray, P.~J.}
\newblock \bibinfo{journal}{\bibinfo{title}{Identifying suitable detection
  dogs}}.
\newblock {\emph{\JournalTitle{Applied Animal Behaviour Science}}}
  \textbf{\bibinfo{volume}{195}}, \bibinfo{pages}{1--7} (\bibinfo{year}{2017}).

\bibitem{troisi2019behavioral}
\bibinfo{author}{Troisi, C.~A.}, \bibinfo{author}{Mills, D.~S.},
  \bibinfo{author}{Wilkinson, A.} \& \bibinfo{author}{Zulch, H.~E.}
\newblock \bibinfo{journal}{\bibinfo{title}{Behavioral and cognitive factors
  that affect the success of scent detection dogs}}.
\newblock {\emph{\JournalTitle{Comparative Cognition \& Behavior Review}}}
  \textbf{\bibinfo{volume}{14}}, \bibinfo{pages}{51--76}
  (\bibinfo{year}{2019}).

\bibitem{lazarowski2020selecting}
\bibinfo{author}{Lazarowski, L.} \emph{et~al.}
\newblock \bibinfo{journal}{\bibinfo{title}{Selecting dogs for explosives
  detection: behavioral characteristics}}.
\newblock {\emph{\JournalTitle{Frontiers in Veterinary Science}}}
  \bibinfo{pages}{597} (\bibinfo{year}{2020}).

\bibitem{ley2008personality}
\bibinfo{author}{Ley, J.}, \bibinfo{author}{Bennett, P.} \&
  \bibinfo{author}{Coleman, G.}
\newblock \bibinfo{journal}{\bibinfo{title}{Personality dimensions that emerge
  in companion canines}}.
\newblock {\emph{\JournalTitle{Applied Animal Behaviour Science}}}
  \textbf{\bibinfo{volume}{110}}, \bibinfo{pages}{305--317}
  (\bibinfo{year}{2008}).

\bibitem{mirko2012preliminary}
\bibinfo{author}{Mirk{\'o}, E.}, \bibinfo{author}{Kubinyi, E.},
  \bibinfo{author}{G{\'a}csi, M.} \& \bibinfo{author}{Mikl{\'o}si, {\'A}.}
\newblock \bibinfo{journal}{\bibinfo{title}{Preliminary analysis of an
  adjective-based dog personality questionnaire developed to measure some
  aspects of personality in the domestic dog (canis familiaris)}}.
\newblock {\emph{\JournalTitle{Applied Animal Behaviour Science}}}
  \textbf{\bibinfo{volume}{138}}, \bibinfo{pages}{88--98}
  (\bibinfo{year}{2012}).

\bibitem{turcsan2018personality}
\bibinfo{author}{Turcs{\'a}n, B.} \emph{et~al.}
\newblock \bibinfo{journal}{\bibinfo{title}{Personality traits in companion
  dogs—results from the vidopet}}.
\newblock {\emph{\JournalTitle{PloS one}}} \textbf{\bibinfo{volume}{13}},
  \bibinfo{pages}{e0195448} (\bibinfo{year}{2018}).

\bibitem{hsu2010factors}
\bibinfo{author}{Hsu, Y.} \& \bibinfo{author}{Sun, L.}
\newblock \bibinfo{journal}{\bibinfo{title}{Factors associated with aggressive
  responses in pet dogs}}.
\newblock {\emph{\JournalTitle{Applied Animal Behaviour Science}}}
  \textbf{\bibinfo{volume}{123}}, \bibinfo{pages}{108--123}
  (\bibinfo{year}{2010}).

\bibitem{serpell2001development}
\bibinfo{author}{Serpell, J.~A.} \& \bibinfo{author}{Hsu, Y.}
\newblock \bibinfo{journal}{\bibinfo{title}{Development and validation of a
  novel method for evaluating behavior and temperament in guide dogs}}.
\newblock {\emph{\JournalTitle{Applied animal behaviour science}}}
  \textbf{\bibinfo{volume}{72}}, \bibinfo{pages}{347--364}
  (\bibinfo{year}{2001}).

\bibitem{van2006phenotyping}
\bibinfo{author}{Van~den Berg, L.}, \bibinfo{author}{Schilder, M.},
  \bibinfo{author}{De~Vries, H.}, \bibinfo{author}{Leegwater, P.} \&
  \bibinfo{author}{Van~Oost, B.}
\newblock \bibinfo{journal}{\bibinfo{title}{Phenotyping of aggressive behavior
  in golden retriever dogs with a questionnaire}}.
\newblock {\emph{\JournalTitle{Behavior genetics}}}
  \textbf{\bibinfo{volume}{36}}, \bibinfo{pages}{882--902}
  (\bibinfo{year}{2006}).

\bibitem{mariti2012perception}
\bibinfo{author}{Mariti, C.} \emph{et~al.}
\newblock \bibinfo{journal}{\bibinfo{title}{Perception of dogs’ stress by
  their owners}}.
\newblock {\emph{\JournalTitle{Journal of Veterinary Behavior}}}
  \textbf{\bibinfo{volume}{7}}, \bibinfo{pages}{213--219}
  (\bibinfo{year}{2012}).

\bibitem{kerswell2009self}
\bibinfo{author}{Kerswell, K.~J.}, \bibinfo{author}{Bennett, P.~J.},
  \bibinfo{author}{Butler, K.~L.} \& \bibinfo{author}{Hemsworth, P.~H.}
\newblock \bibinfo{journal}{\bibinfo{title}{Self-reported comprehension ratings
  of dog behavior by puppy owners}}.
\newblock {\emph{\JournalTitle{Anthrozo{\"o}s}}} \textbf{\bibinfo{volume}{22}},
  \bibinfo{pages}{183--193} (\bibinfo{year}{2009}).

\bibitem{white2011canine}
\bibinfo{author}{White, G.} \emph{et~al.}
\newblock \bibinfo{journal}{\bibinfo{title}{Canine obesity: is there a
  difference between veterinarian and owner perception?}}
\newblock {\emph{\JournalTitle{Journal of Small Animal Practice}}}
  \textbf{\bibinfo{volume}{52}}, \bibinfo{pages}{622--626}
  (\bibinfo{year}{2011}).

\bibitem{rayment2015applied}
\bibinfo{author}{Rayment, D.~J.}, \bibinfo{author}{De~Groef, B.},
  \bibinfo{author}{Peters, R.~A.} \& \bibinfo{author}{Marston, L.~C.}
\newblock \bibinfo{journal}{\bibinfo{title}{Applied personality assessment in
  domestic dogs: Limitations and caveats}}.
\newblock {\emph{\JournalTitle{Applied Animal Behaviour Science}}}
  \textbf{\bibinfo{volume}{163}}, \bibinfo{pages}{1--18}
  (\bibinfo{year}{2015}).

\bibitem{kujala2017human}
\bibinfo{author}{Kujala, M.~V.}, \bibinfo{author}{Somppi, S.},
  \bibinfo{author}{Jokela, M.}, \bibinfo{author}{Vainio, O.} \&
  \bibinfo{author}{Parkkonen, L.}
\newblock \bibinfo{journal}{\bibinfo{title}{Human empathy, personality and
  experience affect the emotion ratings of dog and human facial expressions}}.
\newblock {\emph{\JournalTitle{PloS one}}} \textbf{\bibinfo{volume}{12}},
  \bibinfo{pages}{e0170730} (\bibinfo{year}{2017}).

\bibitem{van2021interspecies}
\bibinfo{author}{van~der Linden, D.}
\newblock \bibinfo{journal}{\bibinfo{title}{Interspecies information systems}}.
\newblock {\emph{\JournalTitle{Requirements Engineering}}}
  \textbf{\bibinfo{volume}{26}}, \bibinfo{pages}{535--556}
  (\bibinfo{year}{2021}).

\bibitem{Riemer2013ChoiceOC}
\bibinfo{author}{Riemer, S.}, \bibinfo{author}{M{\"u}ller, C.~A.},
  \bibinfo{author}{Viranyi, Z.}, \bibinfo{author}{Huber, L.} \&
  \bibinfo{author}{Range, F.}
\newblock \bibinfo{journal}{\bibinfo{title}{Choice of conflict resolution
  strategy is linked to sociability in dog puppies.}}
\newblock {\emph{\JournalTitle{Applied animal behaviour science}}}
  \textbf{\bibinfo{volume}{149 1-4}}, \bibinfo{pages}{36--44}
  (\bibinfo{year}{2013}).

\bibitem{karl2020exploring}
\bibinfo{author}{Karl, S.} \emph{et~al.}
\newblock \bibinfo{journal}{\bibinfo{title}{Exploring the dog--human
  relationship by combining fmri, eye-tracking and behavioural measures}}.
\newblock {\emph{\JournalTitle{Scientific reports}}}
  \textbf{\bibinfo{volume}{10}}, \bibinfo{pages}{1--15} (\bibinfo{year}{2020}).

\bibitem{zamansky2019analysis}
\bibinfo{author}{Zamansky, A.} \emph{et~al.}
\newblock \bibinfo{title}{Analysis of dogs’ sleep patterns using
  convolutional neural networks}.
\newblock In \emph{\bibinfo{booktitle}{International Conference on Artificial
  Neural Networks}}, \bibinfo{pages}{472--483}
  (\bibinfo{organization}{Springer}, \bibinfo{year}{2019}).

\bibitem{bleuer2019computational}
\bibinfo{author}{Bleuer-Elsner, S.} \emph{et~al.}
\newblock \bibinfo{journal}{\bibinfo{title}{Computational analysis of movement
  patterns of dogs with adhd-like behavior}}.
\newblock {\emph{\JournalTitle{Animals}}} \textbf{\bibinfo{volume}{9}},
  \bibinfo{pages}{1140} (\bibinfo{year}{2019}).

\bibitem{fux2021objective}
\bibinfo{author}{Fux, A.} \emph{et~al.}
\newblock \bibinfo{journal}{\bibinfo{title}{Objective video-based assessment of
  adhd-like canine behavior using machine learning}}.
\newblock {\emph{\JournalTitle{Animals}}} \textbf{\bibinfo{volume}{11}},
  \bibinfo{pages}{2806} (\bibinfo{year}{2021}).

\bibitem{menaker2022clustering}
\bibinfo{author}{Menaker, T.}, \bibinfo{author}{Monteny, J.},
  \bibinfo{author}{de~Beeck, L.~O.} \& \bibinfo{author}{Zamansky, A.}
\newblock \bibinfo{journal}{\bibinfo{title}{Clustering for automated
  exploratory pattern discovery in animal behavioral data}}.
\newblock {\emph{\JournalTitle{Frontiers in Veterinary Science}}}
  \textbf{\bibinfo{volume}{9}}, \bibinfo{pages}{884437} (\bibinfo{year}{2022}).

\bibitem{Ren2015FasterRT}
\bibinfo{author}{Ren, S.}, \bibinfo{author}{He, K.}, \bibinfo{author}{Girshick,
  R.~B.} \& \bibinfo{author}{Sun, J.}
\newblock \bibinfo{journal}{\bibinfo{title}{Faster r-cnn: Towards real-time
  object detection with region proposal networks}}.
\newblock {\emph{\JournalTitle{IEEE Transactions on Pattern Analysis and
  Machine Intelligence}}} \textbf{\bibinfo{volume}{39}},
  \bibinfo{pages}{1137--1149} (\bibinfo{year}{2015}).

\bibitem{Lin2014MicrosoftCC}
\bibinfo{author}{Lin, T.-Y.} \emph{et~al.}
\newblock \emph{\bibinfo{title}{Microsoft COCO: Common Objects in Context}}
  (\bibinfo{year}{2014}).

\bibitem{Kuznetsova2018TheOI}
\bibinfo{author}{Kuznetsova, A.} \emph{et~al.}
\newblock \bibinfo{journal}{\bibinfo{title}{The open images dataset v4}}.
\newblock {\emph{\JournalTitle{International Journal of Computer Vision}}}
  \textbf{\bibinfo{volume}{128}}, \bibinfo{pages}{1956--1981}
  (\bibinfo{year}{2018}).

\bibitem{Everingham2010ThePV}
\bibinfo{author}{Everingham, M.}, \bibinfo{author}{Gool, L.~V.},
  \bibinfo{author}{Williams, C. K.~I.}, \bibinfo{author}{Winn, J.~M.} \&
  \bibinfo{author}{Zisserman, A.}
\newblock \bibinfo{journal}{\bibinfo{title}{The pascal visual object classes
  (voc) challenge}}.
\newblock {\emph{\JournalTitle{International Journal of Computer Vision}}}
  \textbf{\bibinfo{volume}{88}}, \bibinfo{pages}{303--338}
  (\bibinfo{year}{2010}).

\bibitem{ts_kmeans}
\bibinfo{author}{Tavenard, R.} \emph{et~al.}
\newblock \bibinfo{journal}{\bibinfo{title}{Tslearn, a machine learning toolkit
  for time series data}}.
\newblock {\emph{\JournalTitle{Journal of Machine Learning Research}}}
  \textbf{\bibinfo{volume}{21}}, \bibinfo{pages}{1--6} (\bibinfo{year}{2020}).

\bibitem{elbow}
\bibinfo{author}{Bholowalia, P.} \& \bibinfo{author}{Kumar, A.}
\newblock \bibinfo{journal}{\bibinfo{title}{Ebk-means: A clustering technique
  based on elbow method and k-means in wsn}}.
\newblock {\emph{\JournalTitle{International Journal of Computer
  Applications}}} \textbf{\bibinfo{volume}{105}}, \bibinfo{pages}{17--24}
  (\bibinfo{year}{2014}).

\bibitem{cnn_ae}
\bibinfo{author}{Yin, C.}, \bibinfo{author}{Zhang, S.}, \bibinfo{author}{Wang,
  J.} \& \bibinfo{author}{Xiong, N.~N.}
\newblock \bibinfo{journal}{\bibinfo{title}{Anomaly detection based on
  convolutional recurrent autoencoder for iot time series}}.
\newblock {\emph{\JournalTitle{IEEE Transactions on Systems, Man, and
  Cybernetics: Systems}}} \textbf{\bibinfo{volume}{52}},
  \bibinfo{pages}{112--122} (\bibinfo{year}{2022}).

\bibitem{adam}
\bibinfo{author}{Zhang, Z.}
\newblock \bibinfo{title}{Improved adam optimizer for deep neural networks}.
\newblock In \emph{\bibinfo{booktitle}{2018 IEEE/ACM 26th International
  Symposium on Quality of Service (IWQoS)}}, \bibinfo{pages}{1--2}
  (\bibinfo{year}{2018}).

\bibitem{grid_search}
\bibinfo{author}{Lerman, P.~M.}
\newblock \bibinfo{journal}{\bibinfo{title}{{Fitting Segmented Regression
  Models by Grid Search}}}.
\newblock {\emph{\JournalTitle{Journal of the Royal Statistical Society Series
  C: Applied Statistics}}} \textbf{\bibinfo{volume}{29}},
  \bibinfo{pages}{77--84} (\bibinfo{year}{2018}).

\bibitem{tpot}
\bibinfo{author}{Olson, R.~S.} \& \bibinfo{author}{Moore, J.~H.}
\newblock \bibinfo{title}{Tpot: A tree-based pipeline optimization tool for
  automating machine learning}.
\newblock In \emph{\bibinfo{booktitle}{Workshop on Automatic Machine
  Learning}}, \bibinfo{pages}{66--74} (\bibinfo{organization}{PMLR},
  \bibinfo{year}{2016}).

\bibitem{teddy_2}
\bibinfo{author}{Lazebnik, T.}, \bibinfo{author}{Somech, A.} \&
  \bibinfo{author}{Weinberg, A.~I.}
\newblock \bibinfo{journal}{\bibinfo{title}{Substrat: A subset-based
  optimization strategy for faster automl}}.
\newblock {\emph{\JournalTitle{Proc. VLDB Endow.}}}
  \textbf{\bibinfo{volume}{16}}, \bibinfo{pages}{772–780}
  (\bibinfo{year}{2022}).

\bibitem{teddy_3}
\bibinfo{author}{Lazebnik, T.}, \bibinfo{author}{Fleischer, T.} \&
  \bibinfo{author}{Yaniv-Rosenfeld, A.}
\newblock \bibinfo{journal}{\bibinfo{title}{Benchmarking biologically-inspired
  automatic machine learning for economic tasks}}.
\newblock {\emph{\JournalTitle{Sustainability}}} \textbf{\bibinfo{volume}{15}},
  \bibinfo{pages}{11232} (\bibinfo{year}{2023}).

\bibitem{teddy_4}
\bibinfo{author}{Keren, L.~S.}, \bibinfo{author}{Liberzon, A.} \&
  \bibinfo{author}{Lazebnik, T.}
\newblock \bibinfo{journal}{\bibinfo{title}{A computational framework for
  physics-informed symbolic regression with straightforward integration of
  domain knowledge}}.
\newblock {\emph{\JournalTitle{Scientific Reports}}}
  \textbf{\bibinfo{volume}{13}}, \bibinfo{pages}{1249} (\bibinfo{year}{2023}).

\bibitem{Frnkranz2010}
\bibinfo{author}{F{\"u}rnkranz, J.} \emph{et~al.}
\newblock In \emph{\bibinfo{booktitle}{Encyclopedia of Machine Learning}}
  (\bibinfo{year}{2010}).

\bibitem{Ling1981CorrelationAC}
\bibinfo{author}{Ling, R.~F.} \& \bibinfo{author}{Kenny, D.~A.}
\newblock \bibinfo{journal}{\bibinfo{title}{Correlation and causation.}}
\newblock {\emph{\JournalTitle{Journal of the American Statistical
  Association}}} \textbf{\bibinfo{volume}{77}}, \bibinfo{pages}{489}
  (\bibinfo{year}{1981}).

\bibitem{teddy_1}
\bibinfo{author}{Savchenko, E.} \& \bibinfo{author}{Lazebnik, T.}
\newblock \bibinfo{journal}{\bibinfo{title}{Computer aided functional style
  identification and correction in modern russian texts}}.
\newblock {\emph{\JournalTitle{Journal of Data, Information and Management}}}
  \textbf{\bibinfo{volume}{4}}, \bibinfo{pages}{25–32}
  (\bibinfo{year}{2022}).

\bibitem{comp_help_1}
\bibinfo{author}{Ramaswamy, S.} \& \bibinfo{author}{DeClerck, N.}
\newblock \bibinfo{journal}{\bibinfo{title}{Customer perception analysis using
  deep learning and nlp}}.
\newblock {\emph{\JournalTitle{Procedia Computer Science}}}
  \textbf{\bibinfo{volume}{140}}, \bibinfo{pages}{170--178}
  (\bibinfo{year}{2018}).

\bibitem{comp_help_2}
\bibinfo{author}{Zanzotto, F.~M.}
\newblock \bibinfo{journal}{\bibinfo{title}{Viewpoint: Human-in-the-loop
  artificial intelligence}}.
\newblock {\emph{\JournalTitle{Journal of Artificial Intelligence Research}}}
  \textbf{\bibinfo{volume}{64}} (\bibinfo{year}{2019}).

\bibitem{comp_help_3}
\bibinfo{author}{Li, G.}
\newblock \bibinfo{journal}{\bibinfo{title}{Human-in-the-loop data
  integration}}.
\newblock {\emph{\JournalTitle{Proceedings of the VLDB Endowment}}}
  \textbf{\bibinfo{volume}{10}}, \bibinfo{pages}{2006--2017}
  (\bibinfo{year}{2017}).

\bibitem{forkosh2021animal}
\bibinfo{author}{Forkosh, O.}
\newblock \bibinfo{journal}{\bibinfo{title}{Animal behavior and animal
  personality from a non-human perspective: Getting help from the machine}}.
\newblock {\emph{\JournalTitle{Patterns}}} \textbf{\bibinfo{volume}{2}},
  \bibinfo{pages}{100194} (\bibinfo{year}{2021}).

\bibitem{menaker2020towards}
\bibinfo{author}{Menaker, T.} \emph{et~al.}
\newblock \bibinfo{title}{Towards a methodology for data-driven automatic
  analysis of animal behavioral patterns}.
\newblock In \emph{\bibinfo{booktitle}{Proceedings of the Seventh International
  Conference on Animal-Computer Interaction}}, \bibinfo{pages}{1--6}
  (\bibinfo{year}{2020}).

\bibitem{volter2023using}
\bibinfo{author}{V{\"o}lter, C.~J.}, \bibinfo{author}{Stari{\'c}, D.} \&
  \bibinfo{author}{Huber, L.}
\newblock \bibinfo{journal}{\bibinfo{title}{Using machine learning to track
  dogs’ exploratory behaviour in the presence and absence of their
  caregiver}}.
\newblock {\emph{\JournalTitle{Animal Behaviour}}}
  \textbf{\bibinfo{volume}{197}}, \bibinfo{pages}{97--111}
  (\bibinfo{year}{2023}).

\end{thebibliography}

\end{document}